\documentclass[12pt]{JHEP3}

\usepackage{amsmath}
\usepackage{amssymb}
\usepackage{graphicx}

\title{Finite size corrections and integrability of
$\mathcal{N}=2$ SYM and DLCQ strings on a pp-wave}

\author{Davide Astolfi\\Dipartimento di Fisica and Sezione I.N.F.N.,
Universit\`a di Perugia, Via A. Pascoli I-06123, Perugia, Italia.
\email{E-mail:astolfi$@$pg.infn.it}}

\author{Valentina Forini\\Dipartimento di Fisica and I.N.F.N. Gruppo
Collegato di Trento, Universit\`a di Trento, 38050 Povo (Trento).
Italia. \email{E-mail:forini@science.unitn.it} }

\author{Gianluca Grignani\\Dipartimento di Fisica and Sezione I.N.F.N.,
Universit\`a di Perugia, Via A. Pascoli I-06123, Perugia, Italia.
\email{E-mail:grignani$@$pg.infn.it}}

\author{Gordon W. Semenoff \\ Department of Physics and Astronomy,
University of British Columbia\\Vancouver, British Columbia, Canada
V6T 1Z1.
\email{E-mail:gordonws$@$phas.ubc.ca}}

\abstract{We compute the planar finite size corrections to the
spectrum of the dilatation operator acting on two-impurity states of
a certain limit of conformal $\mathcal{N}=2$ quiver gauge field
theory which is a $Z_M$-orbifold of $\mathcal{N}=4$ supersymmetric
Yang-Mills theory. We match the result to the string dual, IIB
superstrings propagating on a pp-wave background with a periodically
identified null coordinate. Up to two loops, we show that the
computation of operator dimensions, using an effective Hamiltonian
technique derived from renormalized perturbation theory and a
twisted Bethe ansatz which is a simple generalization of the
Beisert-Dippel-Staudacher~\cite{Beisert:2004hm} long range spin
chain, agree with each other and also agree with a computation of
the analogous quantity in the string theory. We compute the spectrum
at three loop order using the twisted Bethe ansatz and find a
disagreement with the string spectrum very similar to the known one
in the near BMN limit of $\mathcal{N}=4$ super-Yang-Mills theory. We
show that, like in $\mathcal{N}=4$, this disagreement can be
resolved by adding a conjectured ``dressing factor'' to the twisted
Bethe ansatz. Our results are consistent with integrability of the
$\mathcal{N}=2$ theory within the same framework as that of
$\mathcal{N}=4$.}
 \preprint{...}

\begin{document}

\newcommand{\Tr}{\ensuremath{\mathop{\mathrm{Tr}}}}
\newcommand{\sign}{\ensuremath{\mathop{\mathrm{sign}}}}
\newcommand{\del}{\partial}

\def\be{\begin{equation}}
\def\ee{\end{equation}}
\def\bea{\begin{eqnarray}}
\def\eea{\end{eqnarray}}
\def\la{\langle}
\def\ra{\rangle}
\def\dag{\dagger}
\def\wt{\widetilde}
\def\wh{\widehat}

\def\da{\dot{\alpha}}
\def\db{\dot{\beta}}
\def\dg{\dot{\gamma}}
\def\dd{\dot{\delta}}
\def\dl{\dot{\lambda}}
\def\dr{\dot{\rho}}
\def\ds{\dot{\sigma}}
\def\de{\dot{\epsilon}}

\def\G{\Gamma}
\def\D{\Delta}
\def\L{\Lambda}
\def\S{\Sigma}
\def\a{\alpha}
\def\b{\beta}
\def\g{\gamma}
\def\k{\kappa}
\def\d{\delta}
\def\e{\varepsilon}
\def\m{\mu}
\def\tr{\mbox{tr}}
\def\n{\nu}
\def\s{\sigma}
\def\r{\rho}
\def\l{\lambda}
\def\t{\tau}
\def\o{\omega}
\def\O{\Omega}
\def\v{\varrho}
\def\vt{\vartheta}
\def\mc{\mathcal}
\def\N{\nabla}
\def\p{\partial}
\def\K{\widetilde{K}}
\def\tb{\tilde{b}}
\def\lr{\leftrightarrow}
\numberwithin{equation}{section}

\def\journal#1&#2(#3){\unskip, \sl #1\ \bf #2 \rm(19#3) }
\def\andjournal#1&#2(#3){\sl #1~\bf #2 \rm (19#3) }
\def\nextline{\hfil\break}

\def\ie{{\it i.e.}}
\def\eg{{\it e.g.}}
\def\cf{{\it c.f.}}
\def\etal{{\it et.al.}}
\def\etc{{\it etc}}

\def\ap{\alpha'}
\def\sst{\scriptscriptstyle}
\def\tst#1{{\textstyle #1}}
\def\frac#1#2{{#1\over#2}}
\def\coeff#1#2{{\textstyle{#1\over #2}}}
\def\half{\frac12}
\def\hf{{\textstyle\half}}
\def\ket#1{|#1\rangle}
\def\bra#1{\langle#1|}
\def\vev#1{\langle#1\rangle}
\def\d{\partial}

\def\inbar{\,\vrule height1.5ex width.4pt depth0pt}
\def\IC{\relax\hbox{$\inbar\kern-.3em{\rm C}$}}
\def\IR{\relax{\rm I\kern-.18em R}}
\def\IP{\relax{\rm I\kern-.18em P}}
\def\Z{{\bf Z}}
\def\One{{1\hskip -3pt {\rm l}}}
\def\nth{$n^{\rm th}$}
%
%
\def\np#1#2#3{Nucl. Phys. {\bf B#1} (#2) #3}
\def\pl#1#2#3{Phys. Lett. {\bf #1B} (#2) #3}
\def\plb#1#2#3{Phys. Lett. {\bf #1B} (#2) #3}
\def\prl#1#2#3{Phys. Rev. Lett. {\bf #1} (#2) #3}
\def\physrev#1#2#3{Phys. Rev. {\bf D#1} (#2) #3}
\def\prd#1#2#3{Phys. Rev. {\bf D#1} (#2) #3}
\def\prep#1#2#3{Phys. Rep. {\bf #1} (#2) #3}
\def\rmp#1#2#3{Rev. Mod. Phys. {\bf #1} (#2) #3}
\def\cmp#1#2#3{Comm. Math. Phys. {\bf #1} (#2) #3}
\def\cqg#1#2#3{Class. Quant. Grav. {\bf #1} (#2) #3}
\def\mpl#1#2#3{Mod. Phys. Lett. {\bf #1} (#2) #3}

\def\nextline{\hfil\break}
\catcode`\@=11
\def\slash#1{\mathord{\mathpalette\c@ncel{#1}}}
\overfullrule=0pt
\def\AA{{\cal A}}
\def\BB{{\cal B}}
\def\CC{{\cal C}}
\def\DD{{\cal D}}
\def\EE{{\cal E}}
\def\FF{{\cal F}}
\def\GG{{\cal G}}
\def\HH{{\cal H}}
\def\II{{\cal I}}
\def\JJ{{\cal J}}
\def\KK{{\cal K}}
\def\LL{{\cal L}}
\def\MM{{\cal M}}
\def\NN{{\cal N}}
\def\OO{{\cal O}}
\def\PP{{\cal P}}
\def\QQ{{\cal Q}}
\def\RR{{\cal R}}
\def\SS{{\cal S}}
\def\TT{{\cal T}}
\def\UU{{\cal U}}
\def\VV{{\cal V}}
\def\WW{{\cal W}}
\def\XX{{\cal X}}
\def\YY{{\cal Y}}
\def\ZZ{{\cal Z}}
\def\t{\tau}
\def\L{\Lambda}
\def\lam{\lambda}
\def\eps{\epsilon}
\def\vareps{\varepsilon}
\def\underrel#1\over#2{\mathrel{\mathop{\kern\z@#1}\limits_{#2}}}
\def\lapprox{{\underrel{\scriptstyle<}\over\sim}}
\def\lessapprox{{\buildrel{<}\over{\scriptstyle\sim}}}
\catcode`\@=12

\def\sdtimes{\mathbin{\hbox{\hskip2pt\vrule height 4.1pt depth -.3pt width
.25pt \hskip-2pt$\times$}}}
\def\bra#1{\left\langle #1\right|}
\def\ket#1{\left| #1\right\rangle}
\def\vev#1{\left\langle #1 \right\rangle}
\def\det{{\rm det}}
\def\tr{{\rm tr}}
\def\mod{{\rm mod}}
\def \sinh{{\rm sinh}}
\def \cosh{{\rm cosh}}
\def \sgn{{\rm sgn}}
\def\det{{\rm det}}
\def\exp{{\rm exp}}
\def\sh{{\rm sinh}}
\def\ch{{\rm cosh}}

\newcommand{\nn}{\nonumber}
\newcommand{\calO}{{\cal O}}
\newcommand{\spa}{\ \ ,\ \ \ \ }
\newcommand{\bphi}{{\bf \Phi}}

\def\nsp{{NS$^\prime$}}
\def\twoone{{(2,1)}}
\def\twozero{{(2,0)}}
\def\oneone{{(1,1)}}
\def\zeroone{{(0,1)}}
\def\opo{{1+1}}
\def\tpt{{2+2}}
\def\tpo{{2+1}}
\def\j{{\bf j}}
\def\xbar{{\bar x}}
\def\zbar{{\bar z}}
\def\tbar{{\bar\theta}}
\def\psibar{{\bar\psi}}
\def\phibar{{\bar\phi}}
\def\alphabar{{\bar\alpha}}
\def\betabar{{\bar\beta}}
\def\gammabar{{\bar\gamma}}
\def\thetabar{{\bar\theta}}
\def\abar{{\bar a}}
\def\bbar{{\bar b}}
\def\ibar{{\bar i}}
\def\jbar{{\bar j}}
\def\kbar{{\bar k}}
\def\lbar{{\bar \ell}}
\def\mbar{{\bar m}}
\def\dbar{{\bar \d}}

\def\s{{\bf S}}
\def\ij{{i\bar j}}
\def\kahler{{K\"ahler}}
\def\ferm{{\vartheta}}
\def\fermbar{{\bar\vartheta}}
\def\ads{{AdS_3}}
\def\slr{{SL(2)}}
\def\ul{{U(1)}}
\def\nul{{\NN/\ul}}

\def\js{{\bf J}}
\def\gs{{\bf G}}
\def\ls{{\bf L}}

\def\e{\epsilon}

\def\[{[}
\def\]{]}

\def\comment#1{ }

\def\Lc{\hbox{$\cal L$}}
\def\Mc{\hbox{$\cal M$}}
\def\Nc{\hbox{$\cal N$}}

\def\em{\it}
\def\hs{\hskip 5mm}
\def\hsc{\hskip 2mm ,\hskip 5mm}
\def\inv{^{-1}}
\def\pt{\partial}
\def\goto{\rightarrow}
\def\Goto{\Rightarrow}
\def\wbar{\overline}
\def\nl{\item{}}\def\nlb{\item{$\bullet$}}
\def\nlbb{\item{$\bullet\bullet$}}

\def\be{\begin{equation}}
\def\ee{\end{equation}}
\def\bea{\begin{eqnarray}}
\def\eea{\end{eqnarray}}

\section{Introduction}

The idea that the planar limit of $\mathcal{N}=4$ supersymmetric
Yang-Mills theory and its string theory dual, the IIB superstrings
propagating on the AdS$_5\times$S$^5$ background, could both be
exactly integrable has attracted a good deal of
attention~\cite{Beisert:2004hm, Minahan:2002ve, Beisert:2003tq,
Beisert:2003xu, Callan:2003xr, Beisert:2003yb, Serban:2004jf,
Kazakov:2004qf, Callan:2004uv, Arutyunov:2004xr, Beisert:2004jw,
Staudacher:2004tk, Plefka:2005bk, Ambjorn:2005wa, Rej:2005qt,
Frolov:2006cc, Janik:2006dc, Klose:2006dd}. Both ideas have seen
significant development and there is now some hope of an exact
solution of one or both theories.  This could give a remarkably
detailed check of the AdS/CFT correspondence~\cite{Maldacena:1997re,
Gubser:1998bc, Witten:1998qj} at the level of matching planar
Yang-Mills theory to non-interacting strings.

In particular, the gauge theory results have progressed to the point
where integrability has been checked explicitly up to three loop
order~\cite{Serban:2004jf} and there are now proposals for
integrable systems in various sectors of the theory which would be
equivalent to planar Yang-Mills theory to all orders in its loop
expansion~\cite{Serban:2004jf,Beisert:2004hm,Beisert:2005fw,
Beisert:2005tm}.

If string theory on $AdS_5\times S^5$ is integrable, the theory on
simple orbifolds of that space would also be expected to be
integrable.  In the Yang-Mills dual, orbifolding reduces the amount
of supersymmetry and this gives some hope of finding integrability
in theories with less supersymmetry\cite{Kachru:1998ys,
Bertolini:2002nr,Wang:2003cu, DeRisi:2004bc}. In this Paper, we
shall consider the issue of integrability of an $\mathcal{N}=2$
supersymmetric $SU(N)^M$ quiver gauge theory~\cite{Douglas:1996sw}
which can be obtained as a particular $Z_M$-orbifold of
$\mathcal{N}=4$~\cite{Mukhi:2002ck}. This system is also conjectured
to be integrable using a twisted version of the Bethe
ansatz~\cite{Beisert:2005he}. Its string theory dual is IIB
superstrings on the space AdS$_5\times$S$^5/Z_M$.

Thus far, explicit solutions of string theory on these backgrounds
are not known.  Quantitative results are limited to the supergravity
limit, or to some large quantum number
limits~\cite{Alishahiha:2002ev,Kim:2002fp,Takayanagi:2002hv,DeRisi:2004bc}.
For example,  a Penrose limit of AdS$_5\times$S$^5/Z_M$, together
with a large order limit of the orbifold group, $M\to\infty$ can be
taken in such a way that it obtains a
plane-wave~\cite{Metsaev:2001bj} with a periodically identified null
coordinate. The IIB superstring can be solved explicitly in this
background. Mukhi, Rangamani and Verlinde (MRV)~\cite{Mukhi:2002ck}
observed that it is possible to find the Yang-Mills dual of this
theory by taking an analog of the BMN limit~\cite{Berenstein:2002jq,
Kristjansen:2002bb, Beisert:2002bb} of the $\mathcal{N}=2$ quiver
gauge theory. It is a double-scaling limit where $M\to\infty$ and
$N\to\infty$ with the ``effective string coupling'',
$g_2={\small\frac{M}{N}}$, and light-cone radius\footnote{This is
similar to the usual definition of $\lambda'$ in the BMN limit of
$\mathcal{N}=4$ super-Yang-Mills theory,
 $$ \frac{1}{(\alpha'
p^+)^2}=\frac{g_{YM}^2NM}{(kM)^2}\equiv \frac{\lambda'}{k^2} ~~{\rm
or}~~2p^+=\frac{k}{R_-}$$ . }
\begin{equation}R_-={\small\frac{1}{2}}\alpha'\sqrt{g_{YM}^2{\small\frac{N}{M}}} \equiv
{\small\frac{1}{2}}\alpha'\sqrt{\lambda'}\label{rminus}\end{equation}
held finite.

In that limit, they found a beautiful matching of the discrete
light-cone quantized (DLCQ) free string spectrum and planar
conformal dimensions of the appropriate Yang-Mills operators.
Subsequently, some of the simplifying aspects of DLCQ have been
exploited to examine string loop corrections in this
model~\cite{DeRisi:2004bc}.

Our aim in this Paper is to present a computation of the leading
finite size correction to the MRV limit. We will concentrate on
planar Yang-Mills theory and non-interacting strings. In the course
of our work, we will give an explicit demonstration that the twisted
integrability ansatz for the $\mathcal{N}=2$ gauge theory indeed
matches the diagrammatic computation of operator dimensions to two
loop order.

We will compute the $1/M$ corrections to the spectrum of
two-impurity operators to three loop order, ${\lambda'}^3$, in
both the gauge theory and the DLCQ string theory.  We shall find
perfect agreement to two loop order and a disagreement at
three-loop order.

A three-loop order disagreement is already well-known to occur in
the $\mathcal{N}=4$ theory~\cite{Callan:2003xr, Serban:2004jf,
Beisert:2004hm}.  We can check that, in the appropriate limit, our
result matches the one for $\mathcal{N}=4$.

We have tested the statement in Ref.~\cite{Beisert:2005he} that the
orbifolding of $\mathcal{N}=4$ gauge theory results in the
modification of the Bethe ansatz by a simple twist.  Our conclusion
is that it works at least to two-loop order, and we have strong
evidence that it also works at three-loop order. \footnote{An
explicit computation of string energies on orbifolds using twisted
Bethe equation was first considered by
Ideguchi~\cite{Ideguchi:2004wm}.  He computed the spectrum  of
infinite length operators of $\mathcal{N}=0,1,2$ planar orbifold
field theories to one loop order and showed that they matched the
semi-classical spectra of circular string solutions of the strings
in AdS$_5\times$S$^5/Z_M$.}

In addition, we construct the dressing
factor~\cite{Arutyunov:2004xr} that must be taken into account to
find the factorized S-matrix~\cite{Staudacher:2004tk} when the
twisted Bethe ansatz is applied to the string sigma model on the
orbifolded background in the near-MRV limit.

\subsection{Beisert-Dippel-Staudacher ansatz for $\mathcal{N}=4$}

In its most advanced form, the result of integrability of
$\mathcal{N}=4$ super-Yang-Mills theory is a rather simple proposal
for computing dimensions of operators. The typical operators are
composites of the scalar fields $\Phi^i(x)$, $i=1,...,6$.  For
simplicity, we shall concentrate on the $\mathfrak{su}{(2)}$ bosonic
sector.   In that sector, one restricts attention to four of the
scalars in the complex combinations $Z=\Phi^1+i\Phi^2$ and
$\Phi=\Phi^3+i\Phi^4$ and the composite operators
$$
{\rm Tr}\left(\Phi ZZZ\Phi\Phi Z\Phi ZZZ...\right)
$$
At the tree level, since scalar fields have dimension one, the
dimension of this operator is given by the number of scalars that it
contains (we will usually call this $L$).  This spectrum is
degenerate, in that it is the same for whatever scalar fields are
used to make the composite operator. The problem at hand is to
evaluate quantum corrections to the classical dimensions.  These
corrections should resolve the degeneracy. They are obtained by
finding linear combinations of the composite operators which
diagonalize the action of the dilatation operator. The analogy of
this problem with diagonalizing the Hamiltonian of a spin chain, and
the fact that, in the leading order of perturbation theory, it is
identical to the integrable Heisenberg spin chain was observed by
Minahan and Zarembo~\cite{Minahan:2002ve}.

There is a recent proposal which, upon assuming that planar
Yang-Mills theory is integrable, gives an elegant presentation of
the problem of computing operator dimensions to all orders in the
coupling constant~\cite{Beisert:2004hm}. We emphasize at this point,
that we shall only use this proposal up to three loop order, where
its equivalence to renormalized Yang-Mills perturbation theory has
been firmly established. In fact, we shall mainly be interested in a
twisted generalization of it, which is conjectured to describe a
$Z_M$-orbifold of $\mathcal{N}=4$ super-Yang-Mills theory.

In the proposal, the problem for computing eigenvalues of the
dilatation operator is summarized in three equations. First, it
makes use of the Bethe equation for $\mathcal{M}$ magnons on a chain
of length $L$:
\begin{equation} \label{b1}
e^{i p_{j}L}=\prod_{l=1~;~l\neq j}^{\cal M}
\frac{\varphi_{j}-\varphi_{l}+i}{\varphi_{j}-\varphi_{l}-i}=\prod_{l=1~;~l\neq
j}^{\cal M} S(p_j, p_l) \,\;\;\;\;l=1,\dots,{\cal M}
\end{equation}
where $p_i$ are the magnon momenta and $\varphi_i$ are the
corresponding rapidities.  The factorization to 2-body S-matrices
$S(p_i,p_j)$ is also shown.  The momenta in (\ref{b1}) are
constrained by the ``level-matching condition''
\begin{equation}\label{b1a}
\sum_{i=1}^{\mathcal{M}} p_i = 0 ~{\rm mod}~2\pi \end{equation}
which results from the periodicity of the spin chain.  Then, there
is the BDS ``all-loop ansatz''~\cite{Beisert:2004hm}, which are the
remaining two equations. One relates momenta and rapidities, which
depends on the 't hooft coupling $\lambda$,
\begin{equation} \label{b2}
\varphi(p_j)=\frac{1}{2}\cot{\frac{p_j}{2}}\sqrt{1+\frac{\lambda}{
\pi^2}\sin^2{\frac{p_j}{2}}}\ .
\end{equation}
The other gives the spectrum of dimensions as a function of the
momenta,

\begin{equation}\label{b3}
\Delta = L-\mathcal{M}+\sum_{j=1}^{\mathcal{M}} \sqrt{
1+\frac{\lambda}{\pi^2}\sin^2\frac{p_j}{2}} \end{equation} The
program of computing operator dimensions is implemented as follows.
Eqs.~(\ref{b1}) and (\ref{b2}) should first be solved to find $p_i$.
The solutions must depend on $\lambda$ and can in principle be found
at least order-by-order in an expansion in $\lambda$. Then, the
solutions must be inserted into Eq.~(\ref{b3}) to find the operator
dimensions. The statement is that this procedure should yield the
dimensions of this class of operators in $\mathcal{N}=4$
super-Yang-Mills theory. Explicit computations and comparison with
diagrammatic perturbation theory have shown that this procedure
agrees with renormalized  Yang-Mills perturbation theory to at least
third order, and is conjectured to do so for higher orders.  There
is a number of quite non-trivial checks of this fact which are
outlined in Ref.~\cite{Beisert:2004hm}.

\subsection{$\mathcal{N}=2$ quiver gauge theory as orbifolded $\mathcal{N}=4$}

Before we go on to discuss integrability of the $\mathcal{N}=2$
theory, we pause to review some facts about the structure of the
theory and the procedure for computing operator dimensions there.

The $\mathcal{N}=2$ quiver gauge theory with gauge group $SU(N)^M$
is obtained from $\mathcal{N}=4$ with gauge group $SU(MN)$ by a
well-known projection.   Details of this construction can be found
in the
literature~\cite{Douglas:1996sw,Kachru:1998ys,Bershadsky:1998cb}.
The conventions and notation that we use are those of
Refs.~\cite{Mukhi:2002ck},\cite{DeRisi:2004bc} and details can be
found there.

The procedure for obtaining the quiver gauge theory from
$\mathcal{N}=4$ begins by embedding the orbifold group $Z_M$, which
is a subgroup of the R-symmetry group, into the gauge group.  We
will assume that $Z_M$ is in the $\mathfrak{su}{(2)}$ subgroup of
the $\mathfrak{su}{(4)}$ R-symmetry so that orbifolding preserves
$\mathcal{N}=2$ supersymmetry. If $\gamma$ is an element of $Z_M$,
$R(\gamma)$ is the corresponding element of the R-symmetry group and
$U(\gamma)$ is a $U(MN)\times U(MN)$ matrix containing $N$ copies of
the regular representation of $Z_M$, we consider that subset of the
$\mathcal{N}=4$ fields which obey the constraint
\begin{equation} X= U({\gamma})\left[R(\gamma)\circ X
\right] U^\dagger(\gamma)\end{equation} This is accomplished by
setting to zero all of those components which do not obey this
condition. In the present case, choosing $U(\gamma)$ having the
$N\times N$ blocks
$$ U(\gamma)=\left(\begin{matrix}\bar 1 & 0& 0  &
0 & ... \cr
                         0 & \omega & 0& 0 & ... \cr
                         0 & 0  & \omega^2 & 0&...\cr
                         . & . &  . & . & ...\cr
                         0&0&0& ...&\omega^{M-1} \cr
                         \end{matrix}\right)$$
where $\omega=e^{\small \frac{2\pi i}{M}}$ and the action
$$
R(\gamma)Z= \omega Z~~~,~~~R(\gamma)\Phi = \Phi
$$
we see that the surviving components of the two scalar fields which
are of interest to us are $N\times N$ matrices which are embedded in
$MN\times MN$ $\mathcal{N}=4$ variables as follows

\begin{eqnarray}\label{4-2a}
Z = \left(\begin{matrix}0 & 0 & 0 & ...& A_M\cr
                        A_1& 0 & 0 &0&... \cr
                         0 & A_2& 0&0 & ... \cr
                         0 & 0 & A_3&0&...\cr
                         . & . &  . & . & ...\cr
                         0 & 0&0&0& ...\cr
                         \end{matrix}\right)
                         ~~~&,&~~~
\bar Z = \left(\begin{matrix} 0 &\bar A_1& 0  & 0 & ... \cr
                         0 & 0  &\bar A_2& 0 & ... \cr
                         0 & 0  & 0  &\bar A_3&...\cr
                         . & . &  . & . & ...\cr
                         \bar A_M & 0&0&0& ...\cr
                         \end{matrix}\right)
                         \label{trans1}           \\  \label{4-2b}
                         \Phi =\left(\begin{matrix} \Phi_1 & 0& 0  & 0 & ... \cr
                         0 & \Phi_2 & 0& 0 & ... \cr
                         0 & 0  & \Phi_3 & 0&...\cr
                         . & . &  . & . & ...\cr
                         0&0&0& ...&\Phi_M \cr
                         \end{matrix}\right) ~~~&,&~~~
                         \bar \Phi =\left(\begin{matrix}\bar \Phi_1 & 0& 0  & 0 & ... \cr
                         0 & \bar\Phi_2 & 0& 0 & ... \cr
                         0 & 0  & \bar\Phi_3 & 0&...\cr
                         . & . &  . & . & ...\cr
                         0&0&0& ...&\bar\Phi_M \cr
                         \end{matrix}\right)
\label{trans2}                         \end{eqnarray} It is
convenient to think of the blocks as being labelled periodically,
$A_{M+1}=A_1$, etc. The gauge group is $[SU(N)]^M$ with elements
labelled by $U_I$, $I=1,...,M$ and each field transforms as
\begin{eqnarray}\label{gaugetr1}
A_I\to U_IA_IU^\dagger_{I+1} ~~~&,&~~~ \bar A_I\to U_{I+1} A_I
U_I^\dagger
\\ \label{gaugetr2}
\Phi_I\to U_I\Phi_I U_I^\dagger ~~~&,&~~~\bar\Phi_I\to U_I\bar\Phi_I
U_I^\dagger
\end{eqnarray}
States of the $\mathfrak{su}{(2)}$ sector of $\mathcal{N}=4$
super-Yang-Mills were words made from $Z$ and $\Phi$, $$ {\rm
Tr}(ZZ\Phi Z\Phi ZZZZ\Phi ZZZ...)
$$
Since the remaining gauge transformations (\ref{gaugetr1}) and
(\ref{gaugetr2}) now commute with $U(\gamma)$, there are additional
gauge invariant twisted operators
\begin{equation}\label{twistedoperator} {\rm Tr}~\left[U(\gamma)^\ell
ZZ\Phi Z\Phi ZZZZ\Phi ZZZ...\right] ~~,~~ \ell=0,1,...,M-1
\end{equation}
These are translated into words with $(A_I,\Phi_I)$ by substituting
(\ref{trans1}) and (\ref{trans2}). For example,
\begin{equation}\label{vactrace}{\rm Tr}Z^J~~\to~~ M{\rm Tr}
\left[ (A_1A_2...A_M)^k\right]\end{equation} Here, the trace would
vanish unless the total number of fields is given by $J=kM$ with $k$
an integer. In the string theory dual, which is DLCQ strings, the
integer $k$ is the number of units of light-cone momentum and the
operator (\ref{vactrace}) corresponds to the vacuum state of the
string sigma model in the sector with discrete light-cone momentum
$2 p^+=k/R_-$.

States with impurities are made by inserting  $\Phi_I$ into the
trace. Because of the possible twists of the trace, there are more
possible states with these insertions than occurred in the parent
$\mathcal{N}=4$ theory. For example, in $\mathcal{N}=4$, the cyclic
property of the trace implies that there is only one possible
one-impurity state,
$${\rm Tr}\Phi Z^J$$ In the analogous operator of the
$\mathcal{N}=2$ theory, there are $M$ inequivalent one-impurity
states \begin{equation}\label{oneimpurity}{\rm
Tr}\left[A_1...A_{I-1}\Phi_IA_I...A_M(A_1...A_M)^{k-1}\right]
~~~,~~I=1,...,M
\end{equation}
In the string dual,  the extra degrees of freedom that result from
this richer structure turns out to be related to the wrapping number
of the string world sheet on the compact null direction. A naive
Fourier transform of the 1-impurity state, assuming that the are
$kM$ positions that the impurity could take up is
$$\sum_{I=1}^{kM}e^{i{\small\frac{2\pi}{kM}nI}}
{\rm Tr}\left[A_1...A_{I-1}\Phi_IA_I...A_M(A_1...A_M)^{k-1}\right]
~~~,~~n=0,1,...,kM-1
$$
The degree of freedom in the dual string theory corresponding to the
wave-number $n$ in this Fourier transform is the world-sheet
momentum. However, cyclicity of the trace implies that
$n=k\cdot\ell$ where $\ell$ is an integer. This is the
level-matching condition and the integer $\ell$ is dual to the
wrapping number of the string around the periodic null direction.
Once we realize that $n=k\cdot\ell$,  we would recover the twisted
expression (\ref{oneimpurity}), and identify the string wrapping
number $\ell$ with the twist in (\ref{oneimpurity}).

If the orbifold symmetry group is not spontaneously broken, $\ell$
is a good quantum number of the states of the theory and operators
with different values of $\ell$ do not mix with each other.  In
addition it is known that~\cite{Bershadsky:1998cb}, in the planar
limit, the correlation functions of un-twisted operators of the
$\mathcal{N}=2$ theory are identical to those of their parent
operators in $\mathcal{N}=4$ super-Yang-Mills theory once one makes
the replacement $\lambda\to \lambda/M$.  This means that, for the
untwisted operators, with $\ell=0$ in Eq.~(\ref{twistedoperator}),
the dimension should be identical to that in $\mathcal{N}=4$
super-Yang-Mills theory.  This will give a consistency check for
some of our computations in the following.

For the most part, in this Paper we will be interested in
two-impurity operators of the form
\begin{equation}\label{bigoperator}
 \mathcal{O}_{IJ}={\rm Tr}\left(
A_1...A_{I-1}\Phi_IA_I...A_M(A_1....A_M)^pA_1...A_{J-1}\Phi_JA_J...A_M(A_1....A_M)^{k-p-2}\right)
\end{equation}
where we take $I$ and $J$ as running from 1 to $kM$.  Distinct
operators are enumerated by taking $I\leq J$.  The number of scalar
fields in this operator is $kM+2$. The cyclic property of the trace
implies the conditions
\begin{equation}\label{perbc}
\mathcal{O}_{I, kM+1}=O_{1I}
\end{equation}
and
\begin{equation}\label{cyc}
\mathcal{O}_{I+M,J+M}=\mathcal{O}_{I,J}
\end{equation}
which will be important to us.

\subsubsection{The dilatation operator}

Just as in  ${\cal N}=4$ supersymmetric Yang-Mills
theory~\cite{Beisert:2002ff,Beisert:2003tq}, the computation of
dimensions of the operators of interest to us can be elegantly
summarized by the action of an effective Hamiltonian. This technique
was invented in Ref.~\cite{Beisert:2002ff}. The $\mathcal{N}=4$
dilatation operator is known explicitly in terms of its action on
fields up to two loop order, and implicitly to three loop
order~\cite{Beisert:2003tq, Beisert:2003jb, Eden:2004ua}. That part
which is known explicitly can be projected, using the orbifold
projection, to obtain a dilatation operator for the $\mathcal{N}=2$
theory. Here, we shall be interested in computing dimensions of
operators in the scalar $\mathfrak{su}(2)$ sector, so we only retain
the parts of the operator which will contribute there. They can be
obtained by simply substituting the matrices in Eqs.~(\ref{4-2a})
and (\ref{4-2b}) into the analogous terms of the $\mathcal{N}=4$
operator.  The result is

\begin{equation}\label{dil}
D=D_{tree}+D_{1~loop}+D_{2~loops}
\end{equation}
where
\begin{equation}\label{diltree}
 D_{tree}=\sum_{L=1}^M{\rm
Tr}\left(A_L\bar A_L+\Phi_L\bar\Phi_L\right) \end{equation}
\begin{equation}
D_{1~loop}= -\frac{g^{2}_{YM}M}{8\pi^2}\sum_{L=1}^{M}{\rm
Tr}(A_{L}\Phi_{L+1}\bar{A}_{L}\bar{\Phi}_{L}-
A_{L}\Phi_{L+1}\bar{\Phi}_{L+1}\bar{A}_{L} -
\Phi_{L}A_{L}\bar{A}_{L}\bar{\Phi}_{L}+\Phi_{L}A_{L}\bar{\Phi}_{L+1}\bar{A}_{L})
\label{D1}\end{equation}
\begin{eqnarray}
&& D_{2~loops}=\frac{g^{4}_{YM}NM^2}{64\pi^4}\sum_{L=1}^{M}{\rm
Tr}(A_{L}\Phi_{L+1}\bar{A}_{L}\bar{\Phi}_{L}-A_{L}\Phi_{L+1}\bar{\Phi}_{L+1}\bar{A}_L-
\Phi_{L}A_{L}\bar{A}_{L}\bar{\Phi}_{L}+\Phi_{L}A_{L}\bar{\Phi}_{L+1}\bar{A}_{L})
\cr &&+\frac{g^{4}_{YM}M^2}{128\pi^4}\sum_{L=1}^{M}{\rm
Tr}(\Phi_{L}A_{L}\bar{A}_{L}A_{L}\bar{\Phi}_{L+1}\bar{A}_{L}-
A_{L}\Phi_{L+1}\bar{A}_{L}A_{L}\bar{\Phi}_{L+1}\bar{A}_{L}+A_L\Phi_{L+1}A_{L+1}\bar{\Phi}_{L+2}\bar{A}_{L+1}\bar{A}_L
\cr &&- \Phi_{L}A_{L}A_{L+1}\bar{\Phi}_{L+2}\bar{A}_{L+1}\bar{A}_{L}
+ A_{L}\Phi_{L+1}\bar{A}_{L}\bar{\Phi}_{L}\bar{A}_{L-1}A_{L-1}-
\Phi_{L}A_{L}\bar{A}_{L}\bar{\Phi}_{L}\bar{A}_{L-1}A_{L-1} \cr
&&+\Phi_{L}A_{L}\bar{\Phi}_{L+1}\bar{A}_{L}A_{L}\bar{A}_{L}-A_{L}
\Phi_{L+1}\bar{\Phi}_{L+1}\bar{A}_{L}A_{L}\bar{A}_{L} +
A_{L}\Phi_{L+1}\bar{A}_{L}A_{L}\bar{A}_{L}\bar{\Phi}_{L}-
\Phi_{L}A_{L}\bar{A}_{L}\bar{\Phi}_{L}A_{L}\bar{A}_{L}\cr &&-
\Phi_{L}A_{L}\bar{A}_{L}A_{L}\bar{A}_{L}\bar{\Phi}_{L}  +
\Phi_{L}A_{L}A_{L+1}\bar{A}_{L+1}\bar{\Phi}_{L+1}\bar{A}_{L}-
A_{L}\Phi_{L+1}A_{L+1}\bar{A}_{L+1}\bar{\Phi}_{L+1}\bar{A}_{L} \cr
&& + \Phi_{L}A_{L}\bar{A}_{L}\bar{A}_{L-1}\bar{\Phi}_{L-1}A_{L-1}-
A_{L}\Phi_{L+1}\bar{A}_{L}\bar{A}_{L-1}\bar{\Phi}_{L-1}A_{L-1}
+A_{L}\Phi_{L+1}\bar{A}_{L}\bar{\Phi}_{L}A_{L}\bar{A}_{L}) \cr &&
 +\frac{g^{4}_{YM}M^2}{128\pi^4}\sum^{M}_{L=1}{\rm
Tr}(\Phi_{L}A_{L}\bar{\Phi}_{L+1}\Phi_{L+1}\bar{\Phi}_{L+1}\bar{A}_{L}-
A_{L}\Phi_{L+1}\bar{\Phi}_{L+1}\Phi_{L+1}\bar{\Phi}_{L+1}\bar{A}_{L}
+ \Phi_{L}A_{L}\Phi_{L+1}\bar{A}_{L}\bar{\Phi}_{L}\bar{\Phi}_{L} \cr
&&- \Phi_{L}A_{L}\Phi_{L+1}\bar{\Phi}_{L+1}
\bar{A}_{L}\bar{\Phi}_{L} +
A_{L}\Phi_{L+1}\bar{\Phi}_{L+1}\bar{\Phi}_{L+1}\bar{A}_{L}\Phi_{L}
-\Phi_{L} A_{L}\bar{\Phi}_{L+1}\bar{\Phi}_{L+1}\bar{A}_{L}\Phi_{L}
\cr &&-
A_{L}\Phi_{L+1}\bar{\Phi}_{L+1}\bar{A}_{L}\Phi_{L}\bar{\Phi}_{L}
+A_{L}\Phi_{L+1}\bar{\Phi}_{L+1}\Phi_{L+1}
\bar{A}_{L}\bar{\Phi}_{L}+\Phi_L
A_L\bar{\Phi}_{L+1}\bar{A}_L\Phi_L\bar{\Phi}_L\cr
&&-\Phi_{L}A_{L}\bar{\Phi}_{L+1}\Phi_{L+1}\bar{A}_{L}\bar{\Phi}_{L}
+A_{L}\Phi_{L+1}\Phi_{L+1}\bar{\Phi}_{L+1}\bar{A}_{L}\bar{\Phi}_{L}-A_{L}\Phi_{L+1}\Phi_{L+1}
\bar{A}_{L}\bar{\Phi}_{L}\bar{\Phi}_{L} \cr &&+
\Phi_{L}A_{L}\bar{\Phi}_{L+1}\bar{A}_{L}\bar{\Phi}_{L}\Phi_{L}-
A_{L}\Phi_{L+1}\bar{\Phi}_{L+1}\bar{A}_{L}\bar{\Phi}_{L}\Phi_{L} +
A_{L}\Phi_{L+1 }\bar{A}_{L}\bar{\Phi}_{L}\Phi_{L}\bar{\Phi}_{L}-
\Phi_{L}A_{L}\bar{A}_{L}\bar{\Phi}_{L}\Phi_{L}\bar{\Phi}_{L} )\cr&&
\label{d2}\end{eqnarray}    The number of loops which contribute to
each order is exhibited in the power of the Yang-Mills coupling
constant $g^2_{YM}$ which precedes each term. Later  we will use
either the parent $\mathcal{N}=4$ 't hooft coupling,
$$\lambda\equiv g_{YM}^2NM$$ which is important for the planar
limit,  or the modified 't hooft coupling
$$\lambda'\equiv \frac{g_{YM}^2N}{M}=\frac{\lambda}{M^2}$$ which is
held constant in the MRV limit.   In the latter limit, $N$ and $M$
are both put to infinity so that $\lambda'$ and the effective string
coupling, $$g_2\equiv \frac{M}{N}$$ are held constant.   The
effective string coupling controls the appearance of non-planar
diagrams and, to get the planar limit, which we will for the most
part be interested in, it must also be put to zero. Inspection of
the 1-loop and 2-loop dilatation operators shows that, in order for
this MRV limit to make sense, their action should be suppressed by
some powers of $\frac{1}{M}$ further to those exhibited in
Eqs.~(\ref{D1}) and (\ref{d2}). We shall see that this is indeed the
case.

The action of the operators in Eqs.~(\ref{diltree}), (\ref{D1}) and
(\ref{d2}) on a composite of the form (\ref{bigoperator}) is
implemented with the following procedure.

We note that each term in the dilatation operators contains a few
$\bar A_I$'s and $\bar\Phi_I$'s. We take a term in $D$, and we
Wick-contract the $\bar A_I$'s and $\bar\Phi_I$'s which appear in
that term with each occurrence of $A_I$ and $\Phi_I$ in the trace
(\ref{bigoperator}) according to the rules

$$
\left<\left[\bar A_I\right]_{ab}
\left[A_{J}\right]_{cd}\right>_0=\delta_{IJ}\delta_{ad}\delta_{bc}
~~,~~ \left<\left[\bar \Phi_I\right]_{ab}
\left[\Phi_{J}\right]_{cd}\right>_0=\delta_{IJ}\delta_{ad}\delta_{bc}
$$
Here we are treating the fields as if they are simply matrices in a
Gaussian matrix model, ignoring their space-time dependence and
simply substituting them with other fields according to the rules of
performing the contractions. The space-time dependence, that of
course must be taken into account in order to compute dimensions in
renormalized perturbation theory, has already been taken care of in
formulating the effective Hamiltonian.

In doing these contractions with the first term in (\ref{dil}), the
tree-level operator, we find the tree level contribution to the
conformal dimension.  The procedure merely counts the number of
scalar fields, giving $kM+2$ in the case of (\ref{bigoperator}).

When we Wick-contract with the 1-loop and 2-loop terms, (\ref{D1})
and (\ref{d2}), once all possible contractions are done, we find a
superposition of operators where the total number of fields in each
operator is the same and the number of impurities in each operator
is still two, but the positions of the impurities have been shifted.

All of the operators in the superposition have the same tree-level
dimensions. It means that, at the outset, we could have began with
linear combinations of them.  We could then have chosen the
coefficients in the linear combinations in such a way as to
diagonalize the action of the dilatation operator.  Upon doing this,
we would find the eigenvalues, i.e. the dimensions,  and the linear
combinations that we find would be the scaling operators themselves.

Once the Wick contractions are explicitly performed, the action of
the one loop dilatation operator on the operators
(\ref{bigoperator}) is given by two equations, depending on whether
the impurities lie next to each other or not
\begin{equation}\label{D1IJ}
D_{1~loop}\circ O_{IJ}=\frac{\lambda' M^2 }{8 \pi^2}\big(-O_{I+1,
J}-O_{I-1, J}+4O_{IJ}-O_{I,J+1}-O_{I,J-1}\big),\quad I<J
\end{equation}
\begin{equation}\label{D1II}
D_{1~loop}\circ O_{II}=\frac{\lambda' M^2}{8 \pi^2}\big(-O_{I-1,
I}-O_{I,I+1}+2O_{II}\big)
\end{equation}
At two loops, the action of the dilation operator results in three
equations,
\begin{eqnarray}\label{D2IJ}
D_{2~loops}\circ O_{IJ}&=&\frac{\lambda'^2 M^4 }{128
\pi^4}\big(-O_{I-2, J}-O_{I+2, J}+4O_{I-1, J}+4O_{I+1, J} \cr &-&
O_{I, J-2}-O_{I, J+2}+4O_{I, J-1}+4O_{I, J+1}-12O_{IJ}\big)
\end{eqnarray}
for $J-I\geq 2$ and
\begin{eqnarray}\label{D2II}
D_{2~loops}\circ O_{II}&=&\frac{ \lambda'^2 M^4 }{128
\pi^4}\big(-O_{I-2, I}+4O_{I-1, I}-O_{I-1, I-1} \cr &-& 4 O_{I,
I}+4O_{I, I+1}-O_{I+1, I+1}-O_{I, I+2}\big)
\end{eqnarray}
\begin{eqnarray}\label{D2II+1}
D_{2~loops}\circ O_{I,I+1}&=&\frac{\lambda'^2 M^4 }{128
\pi^4}\big(-O_{I, I+3}+4 O_{I+1, I+1}+4O_{I, I+2}-14 O_{I, I+1} \cr
&+& 4 O_{I, I}+4O_{I-1, I+1}-O_{I-2, I+1}\big)
\end{eqnarray}
where the second and the third formulae represent, respectively, the
nearest ($I=J$) and the next-to-nearest ($J=I+1$) neighbor cases. We
see explicitly that the dilatation operator is acting like a lattice
differential operator on the matrix chains. The result is an
effective spin-chain Hamiltonian. The problem of finding the
eigenvalues of this Hamiltonian is integrable and can be attacked
using the twisted Bethe ansatz, which we summarize in the next
subsection.

\subsection{Twisted Bethe ansatz for the orbifold}

The conjecture~\cite{Beisert:2005he} is that the spectrum of
operator dimensions in the $\mathfrak{su}(2)$ sector of the
$\mathcal{N}=2$ quiver theory which is a $Z_M$ orbifold of
$\mathcal{N}=4$ is found by including a simple twist in the Bethe
equation (\ref{b1}).  The other equations, (\ref{b2}) and (\ref{b3})
are applied unchanged.

For example, for two magnons, the twisted Bethe equations are
\begin{equation} \label{tb1}
e^{i p_{1}(kM+2)}=\omega^\ell~
\frac{\varphi_{1}-\varphi_{2}+i}{\varphi_{1}-\varphi_{2}-i}
~~,~~e^{i p_{2}(kM+2)}=\omega^\ell~
\frac{\varphi_{2}-\varphi_{1}+i}{\varphi_{2}-\varphi_{1}-i}
\end{equation}
Here, as in (\ref{b1}), $L=kM+2$ is the length of the chain. The
twist is the $M$'th root of unity factor $\omega^\ell$ in front the
right-hand-sides of (\ref{tb1}).  $\omega=e^{\frac{2\pi}{M}i}$ and
the integer $\ell$ is the charge of the state under the U(1)
symmetry which is used in the orbifold projection.  In the dual
string theory, it coincides with the wrapping number of the string
world-sheet on the compact null direction. Because of (\ref{tb1a}),
it is related to the total world-sheet momentum
$e^{i(p_1+p_2)}=\omega^\ell$. As in the $\mathcal{N}=4$ theory, the
momenta and rapidities are still related by
\begin{equation} \label{tb2}
\varphi_1=\frac{1}{2}\cot{\frac{p_1}{2}}\sqrt{1+\frac{\lambda}{
\pi^2}\sin^2{\frac{p_1}{2}}}~~,~~\varphi_2=
\frac{1}{2}\cot{\frac{p_2}{2}}\sqrt{1+\frac{\lambda}{
\pi^2}\sin^2{\frac{p_2}{2}}}\ .
\end{equation}
and the spectrum is

\begin{equation}\label{tb3}
\Delta = kM+ \sqrt{ 1+\frac{\lambda}{\pi^2}\sin^2\frac{p_1}{2}} +
\sqrt{ 1+\frac{\lambda}{\pi^2}\sin^2\frac{p_2}{2}}
\end{equation}

Multiplying the two equations in (\ref{tb1}) gives the condition on
the total momentum
\begin{equation}
e^{i(p_1+p_2)kM}=1~~\to ~~ p_1+p_2 = \frac{2\pi}{kM}s ~~,~~~s={\rm
integer}
\end{equation}
The ``level-matching condition'' (\ref{b1a}) is replaced by
\begin{equation}\label{tb1a} \sum_{i=1}^{\cal{M}} p_i =
\frac{2\pi}{M}\cdot\ell ~~~,~~\ell=~{\rm integer}\end{equation} and
it  implies
\begin{equation}
s~=~k\cdot~{\rm integer}
\end{equation}

It is clear from the form of the equations (\ref{tb1}) and
(\ref{tb2}) that the momenta, which are their solutions, generally
depend on $\lambda$ and the parameter $kM$.  It is also clear that
the momenta which solve them must be small when $M$ is large,
$p_i\propto \frac{1}{kM}$.  This is also needed for consistency of
the MRV limit where $M\to\infty$ and $\lambda\to\infty$ in such a
way that $\lambda'=\frac{\lambda}{M^2}$ remains finite.  Equation
(\ref{tb2}) also implies that $\varphi_1$ and $\varphi_2$ are both
of order $M$ in that limit.  Later in this Paper, we shall consider
the leading corrections to this limit in an expansion in $1/M$.  In
the remainder of this subsection, for a warmup exercise, we will
seek the solutions for $p_i$ in the MRV limit, where $M\to\infty$.
In this limit, we hold $\lambda'={\small\frac{\lambda}{M^2}}$
finite.

Even in this limit, we shall not be able to solve  equations
(\ref{tb1}) and (\ref{tb2}) for arbitrary values of $\lambda'$. We
will be limited to considering a Taylor expansion of Eq.~(\ref{tb2})
in $\lambda'$ and then seeking momenta which are also expressed as
expansions in $\lambda'$. We begin with the leading order where we
simply set $\lambda'$ to zero in Eq.~(\ref{tb2}). \footnote{We do
this by setting $\lambda$ to zero, but we must be careful to see, a
posteriori, that indeed
$p_i\sim\mathcal{O}\left({\small\frac{1}{M}}\right)$, so that
setting $\lambda=0$ is equivelent to setting $\lambda'=0$. We shall
see this shortly, in Eq.~(\ref{mom0}).} Then, it is easy to see that
the momenta must be given by

\begin{equation}\label{mom0}p_1=\frac{2\pi}{kM}n_1
+\mathcal{O}\left(\frac{1}{M^2}\right)~~, ~~ p_2=\frac{2\pi}{kM}n_2
+\mathcal{O}\left(\frac{1}{M^2}\right)\end{equation} where $n_1$ and
$n_2$ are integers.  Level matching gives the further condition
$$n_1+n_2=k\cdot\ell$$ where $\ell$ is an integer. Then
Eq.~(\ref{tb3}) implies
\begin{equation}
\Delta = kM+\sqrt{ 1+\lambda'\small{\frac{n_1^2}{k^2}} }
+\sqrt{1+\lambda'{\small\frac{n_2^2}{k^2}} }
\end{equation}
which agrees beautifully with the spectrum of DLCQ free strings on
the plane-wave background.

\subsection{Coordinate Bethe ansatz}
\label{sec:coordinate}

There is another, equivalent procedure which is sometimes
convenient, called the coordinate Bethe ansatz. Since we will make
use of it later, we shall review it here for the special case of a
two-impurity operator.

Consider the dilatation operator in the form of the difference
operators (\ref{D1IJ})-(\ref{D2II+1}) which we derived using the
effective Hamiltonian. Finding the spectrum of the dilatation
operator entails finding the eigenstates and eigenvalues of the
combination of difference operators (\ref{D1IJ})-(\ref{D2II+1}),
operating on the space of two-impurity operators. Here, for
illustration, we will review the argument that, to order $\lambda'$,
this is equivalent to the task of solving the twisted Bethe ansatz
which was set out in the previous sub-section.  Later on in this
Paper, we will show that this also holds to order ${\lambda'}^2$
(and then we will assume that it holds to order ${\lambda'}^3$).

To begin, we take the linear super-position of two-impurity
operators
\begin{equation}
\mathcal{O}\equiv \sum_{1\leq I\leq J\leq kM}\Psi_{IJ}O_{IJ}
\label{eigenst}
\end{equation}
Our task is to find the coefficients $\Psi_{IJ}$ in this series so
that this operator is an eigenstate of the dilation operator. If we
impose the same periodicity conditions on $\Psi_{IJ}$ as the
operators $O_{IJ}$ obey in (\ref{perbc}), the action of the
dilatation operator as  difference operators in
(\ref{D1IJ})-(\ref{D2II+1}) is self-adjoint\footnote{We note that
the detailed form of the contact terms in the difference operators
are essential in demonstrating the self-adjoint property.} and we
can recast the problem of diagonalizing dilatations as the problem
of finding eigenvalues for the action of the difference operators
acting on the wave-functions $\Psi_{IJ}$.

The coordinate Bethe ansatz was used in refs.~\cite{Ideguchi:2004wm}
and \cite{DeRisi:2004bc} to find the spectrum of the one-loop
operator in the large $M$ limit. To introduce the technique, we
shall review the essential parts of the argument here.  At one-loop
order, the eigenvalue equation is
\begin{eqnarray}\label{schr1}
E^{(1)}\Psi_{IJ}&=&g^2\left(-\Psi_{I+1, J}-\Psi_{I-1,
J}+4\Psi_{IJ}-\Psi_{I,J+1}-\Psi_{I,J-1}\right)~~~~I<J \\
E^{(1)}\Psi_{IJ}&=&g^2\left(-\Psi_{I-1,
I}-\Psi_{I,I+1}+2\Psi_{II}\right)~~~~~~~~~~~~~~~~I=J \label{schr2}
\end{eqnarray}
where $g^2=g_{YM}^2 N M/(8\pi^2)$.   To look for a solution, we make
the plane-wave ansatz
\begin{equation}
\Psi_{IJ}=\mu_1^I\mu_2^J+S_0(\m_2, \m_1)\mu_2^{I}\mu_1^{J}
\label{ans1}
\end{equation}
where $\mu_1=e^{i p_1}$ and $\mu_2=e^{i p_2}$. Then,
Eq.~(\ref{schr1}) yields the eigenvalue,
\begin{equation}
E^{(1)}=\frac{\lambda' M^2}{2\pi^2}\left(\sin^2 \frac{p_1}{2}+\sin^2
\frac{p_2}{2}\right) \label{eigenv1}
\end{equation}
which is the expansion to first order in $\lambda'$ of the square
roots in (\ref{tb3}).  The problem of finding the allowed values
of $(p_1,p_2)$ remains.

Then, (\ref{schr2}) yields the equation

\begin{equation} \label{S0}
S_0(\m_2, \m_1)=-\frac{\mu_1}{\mu_2}
\frac{\mu_1\mu_2-2\mu_2+1}{\mu_1\mu_2-2\mu_1+1}
\end{equation}
where it should be noticed that
$S_0(\m_1,\m_2)^{-1}=S_0(\m_2,\m_1)$.

The boundary condition $\Psi_{I,kM+1}=\Psi_{1,I}$ gives
\begin{equation}
\m_2^{kM}=S_0(\m_2,\m_1)~~~,~~~\m_1^{kM}=S_0(\m_2,\m_1)^{-1}
\label{bcoord1}
\end{equation}

Eqs. (\ref{bcoord1}) together with (\ref{S0}) are identical to the
twisted Bethe equations (\ref{tb1}), together with (\ref{tb2})
with $\lambda'$ set to zero. The level-matching condition is
obtained by noticing that
\begin{eqnarray} \label{boundary3}
& \Psi_{I+M, J+M}=\Psi_{IJ}
\end{eqnarray}
implies
\begin{eqnarray} \label{boundary4}
  \left(\mu_1\mu_2\right)^{M}=1
\end{eqnarray}

\subsection{Outline}

In the remainder of this Paper, we shall compute the finite size
corrections to the spectrum of dimensions of the two-impurity
operators in the $\mathfrak{su}(2)$ bosonic sector that we have been
discussing so far.  We will use the twisted Bethe ansatz, summarized
in Eqs.~(\ref{tb1})-(\ref{tb3}) and will compute to three-loop
order. We also will check explicitly that the coordinate Bethe
ansatz technique which used the difference operator form of the
dilatation operator exhibited in Eqs.~(\ref{D1IJ})-(\ref{D2II+1})
indeed produces the same result to two loop order.

Then, we will adopt the string theory computation which was
originally used in Ref.~\cite{Callan:2003xr} for the near pp-wave
limit of $AdS_5\times S^5$ to the present case of the near DLCQ
pp-wave limit of $AdS_5\times S^5/Z_M$. This is the string theory
dual of the ``near''-MRV limit of the $\mathcal{N}=2$ theory.  We
compute the spectrum of the string in this case, expanded to order
$1/M$. On the string side, the expression that is obtained is exact
to all orders in $\lambda'$.  When expanded to third order, we find
beautiful agreement with the $\mathcal{N}=2$ gauge theory prediction
up to second order in $\lambda'$, i.e. two loops, and disagreement
at third, or three loop order.

This disagreement is similar to the one which is found in the
$\mathcal{N}=4$ theory in Ref.~\cite{Serban:2004jf, Beisert:2004hm}.
In fact, in the de-compactified limit, $k\to\infty,R_-\to\infty$
with $p^+=k/R_-$ fixed, it approaches that result.

In addition, we show that, like in the case of $\mathcal{N}=4$
super-Yang-Mills theory, the discrepancy can be taken into account
by a dressing factor~\cite{Staudacher:2004tk}.

\section{Finite size corrections at one loop}

In order to calculate the first finite size corrections to
Eq.(\ref{mom0}) we make the following general ansatz for the magnon
momenta
\begin{eqnarray}\label{oneloopansatz} p_1=\frac{2n_1\pi}{k M}+\frac{A\pi}{M^2} \cr
p_2=\frac{2n_2\pi}{k M}-\frac{A\pi}{M^2}
\end{eqnarray}
Recall that we solve at one loop order by simply setting
$\lambda'\to 0$ in the equation for the rapidity (\ref{tb2}), so
that it is given by
\begin{equation}\label{rapidityoneloop}
\varphi_j=\frac{1}{2}\cot{\frac{p_j}{2}}.
\end{equation}
By requiring that the Bethe equations (\ref{tb1}) are satisfied by
(\ref{oneloopansatz}) at both leading and next to leading order in
$\frac{1}{M}$ one gets the following value for $A$
\begin{equation}
\label{A} A=\frac{2\left(n_1^2+n_2^2\right)}{k^2(n_2-n_1)}
\end{equation}
We can then insert this solution in the expression (\ref{eigenv1})
for the anomalous dimension in terms of $p_i$ and expand in a
$\frac{1}{M}$ series. The first finite size correction to the planar
anomalous dimension reads
\begin{equation}\label{finitesizeoneloop}
\Delta_{1~loop}=\frac{\lambda'}{2}\left[
\frac{n_1^2+n_2^2}{k^2}-\left(\frac{2}{kM}\right)
\frac{\left(n_1^2+n_2^2\right)}{k^2}+O\left(\frac{1}{M^2}\right)\right]
\end{equation}
 As a first
consistency check, it is easy to verify that when the
$\mathcal{N}=4$ level-matching condition $n_2=-n_1$ is imposed --
this gives the result for the unwrapped, $\ell=0$ state -- recalling
that $J=kM$ and the appropriate re-definition of $\lambda'$, the
$\mathcal{N}=4$ result~\cite{Serban:2004jf, Beisert:2004hm} is
recovered.

The zeroth order term in (\ref{finitesizeoneloop}) equals the
one-loop free string spectrum in the plane-wave limit and the first
finite size correction, $\frac{1}{M}$ order, will be compared with
the corresponding $1/R^2$ correction on the string side of the
duality.

\section{Two loops}

To find the correction to the dimension at two loops, we must
expand (\ref{tb2}) to linear order in $\lambda'$ and then use it
in (\ref{tb1}) to find the momenta, also to linear order in
$\lambda'$. The resulting  twisted Bethe equation reads
\begin{equation}\label{bethetwoloops}
e^{i
p_2(kM+2)}=e^{i(p_1+p_2)}\frac{\frac{1}{2}\cot\frac{p_2}{2}+\frac{\l}{8\pi^2}\sin
p_2-\frac{1}{2}\cot\frac{p_1}{2}+\frac{\l}{8\pi^2}\sin
p_1+i}{\frac{1}{2}\cot\frac{p_2}{2}+\frac{\l}{8\pi^2}\sin
p_2-\frac{1}{2}\cot\frac{p_1}{2}+\frac{\l}{8\pi^2}\sin p_1-i}
\end{equation}

The simultaneous expansion of the momenta in $\lambda'$ and
$\frac{1}{M}$ will have the form
\begin{eqnarray}
\label{twoloopsansatz}
p_1=\frac{2n_1\pi}{kM}+\frac{A\pi}{M^2}+\lambda'\frac{B\pi}{M^2}+...
~~,~~
p_2=\frac{2n_2\pi}{kM}-\frac{A\pi}{M^2}-\lambda'\frac{B\pi}{M^2}+...
\end{eqnarray}
where $A$, given in Eq.~(\ref{A}), was calculated in the previous
section. We could also have included a contribute of order
$\lambda'/M$ to the momenta, but Eq.(\ref{bethetwoloops}), expanded
as a power series in $\l'$ and $1/M$, would force it to be zero.

The corrections, indicated by three dots are at least of order
$\frac{1}{M^3}$ or $\frac{{\lambda'}^2}{M^2}$. (In the next Section,
we will compute the $\frac{{\lambda'}^2}{M^2}$ correction.)

$B$ can be fixed by requiring that the Bethe equation
(\ref{bethetwoloops}) is satisfied at the first order in the
$\lambda'$ expansion
\begin{equation}
\label{B} B=\frac{2~n_1^2 n_2^2}{k^4(n_2-n_1)}
\end{equation}

To calculate the $O(\lambda'^2)$ contribution to the planar
anomalous dimension, one plugs the solution of the Bethe equation
into the eigenvalue formula (\ref{tb3}). Performing a double series
expansion, in $\lambda'$ and $\frac{1}{M}$, we obtain the following
expression for the two loops planar anomalous dimension, up to the
first finite size correction
\begin{equation}\label{twoloopsanomalousdimension}
\Delta_{2~loops}=\frac{\lambda'^2}{8}\left[-\frac{n_1^4+n_2^4}{k^4}+
\left(\frac{4}{k M}\right)\frac{n_1^4+n_1^3 n_2+n_1
n_2^3+n_2^4}{k^4}+O\left(\frac{1}{M^2}\right)\right].
\end{equation}
As a consistency check, we take the case where  $\ell=(n_1+n_2)/k=0$
We see that (\ref{twoloopsanomalousdimension}) agrees with the
$\mathcal{N}=4$ solution~\cite{Serban:2004jf, Beisert:2004hm} in
that case.

\section{Two loops revisited: the perturbative asymptotic Bethe
ansatz}

In order to diagonalize the two-loop corrected dilatation operator
(\ref{dil}) the ansatz for the wave-function (\ref{ans1}) has to be
adjusted in a perturbative sense in order to take into account long
range interactions. When interactions are included at the next
order, the wave-functions are no longer plane waves.  The technique
which is used, termed as perturbative asymptotic Bethe ansatz
(PABA)~\cite{Klose:2003qc, Staudacher:2004tk}, begins with
\begin{equation}
\Psi_{IJ}=\m_1^I\m_2^J~ f(J-I+1,\m_1,\m_2)+ \m_2^I\m_1^J~ f(k
M-J+I+1,\m_1,\m_2)~S(\m_2,\m_1)\label{PABA}
\end{equation}
where the $S$-matrix and the function $f$ have the perturbative
expansions
\begin{eqnarray}
S(\m_2,\m_1)&=&S_0(\m_2,\m_1)+\sum_{n=1}^\infty (g^2)^n
~S_n(\m_2,\m_1)\cr f(J-I+1,\m_1,\m_2)&=&1+\sum_{n=0}^\infty
(g^2)^{n+|J-I+1|}f_n(\m_1,\m_2) \label{expans}
\end{eqnarray}
where $g^2=g^2_{\rm{YM}}M N/(8\pi^2)=\lambda' M^2/(8\pi^2)$. The
number of powers of the coupling in the second of Eqs.(\ref{expans})
clearly indicates the interaction range on the lattice.

Note that, once it is determined at the leading order, the
wave-function at the next order should be uniquely determined by
quantum mechanical perturbation theory. Here, we are postulating
that the result of determining it can be put in the form of
Eq.~(\ref{PABA}).  We will justify this postulate by showing that
(\ref{twoloopsanomalousdimension}) does satisfy the equation to the
required order and that the process of finding the solution is
encoded in the twisted Bethe ansatz.

To derive the two loop Bethe equations it is sufficient to keep only
the following terms in the ansatz (\ref{PABA})
\begin{eqnarray}
\Psi_{IJ}&=&\m_1^I\m_2^J\left[1+g^{2|J-I+1|}f_0(\mu_1,\mu_2)\right]\cr&+&\m_2^I\m_1^J\left[S_0(\m_2,\m_1)+g^2S_1(\m_2,\m_1)\right]\left[1+g^{2|k
M+1-J+I|}f_0(\mu_1,\mu_2)\right] \label{2loopsansatz}
\end{eqnarray}
The boundary conditions $\Psi_{I,kM+1}=\Psi_{1,I}$ on
(\ref{2loopsansatz}) imply the Bethe equations
\begin{eqnarray} \label{bethe2}
& \mu_2^{kM}=S_0(\m_2,\m_1)+g^2 S_1(\m_2,\m_1) \cr &
\m_1^{kM}=[S_0(\m_2,\m_1)+g^2 S_1(\m_2,\m_1)]^{-1}
\end{eqnarray}

The Schr\"{o}dinger equation is obtained, as in Section
\ref{sec:coordinate}, by acting on the wave-function $\Psi_{IJ}$
with the dilatation operator as difference operators according to
(\ref{D1IJ})-(\ref{D2II+1}). In doing so, the two-loop contributions
coming from the action of the 1-loop dilatation operator on the
order $\l'$ part of the wave-function have to be kept into account.
Note that, since $\mu_i=e^{i p_i}$ and in general the $p_i$'s depend
on $\l'$, the wave function has an implicit dependence on $\l'$
through its dependence on $\mu_i$.

The difference equation for $J-I\geq 2$ reads
\begin{eqnarray}
&&\left(D_{1~loop}+D_{2~loop}\right)\circ\Psi_{IJ}=\cr
&&g^2\left(-\Psi_{I+1, J}-\Psi_{I-1,
J}+4\Psi_{IJ}-\Psi_{I,J+1}-\Psi_{I,J-1}\right) \cr &&\frac{g^4
}{2}\left(-\Psi_{I-2, J}-\Psi_{I+2, J}+4\Psi_{I-1, J}+4\Psi_{I+1, J}
\right.\cr &&\left.- \Psi_{I, J-2}-\Psi_{I, J+2}+4\Psi_{I,
J-1}+4\Psi_{I, J+1}-12\Psi_{IJ}\right)~~~~~~~~J-I\geq 2
\end{eqnarray}
Using the ansatz (\ref{2loopsansatz}) and keeping only terms up to
order $g^4$ we see that, when $J-I\geq 2$  the dilatation operator
acting on the wave-function returns its form times an eigenvalue,

\begin{equation}
\left(D_{1~loop}+D_{2~loop}\right)\circ\Psi_{IJ}=\left[4
g^2\left(\sin^2\frac{p_1}{2}+\sin^2\frac{p_2}{2}\right)-\frac{g^4}{8}\left(\sin^4\frac{p_1}{2}+\sin^4\frac{p_2}{2}\right)\right]\Psi_{IJ}
\end{equation}
In order for (\ref{2loopsansatz}) to be a  eigenstate of the
dilatation operator up to two loops, this must also be so for the
contact terms in the dilatation operator.  For this, the following
equations must hold:

\begin{eqnarray}
&&\left(D_{1~loop}+D_{2~loop}\right)\circ\Psi_{II}=\cr
&&g^2\left(-\Psi_{I-1,I}-\Psi_{I,I+1}+2 \Psi_{I,I}\right)\cr
&&+\frac{g^4}{2}\left(-\Psi_{I-2,I}+4\Psi_{I-1,I}-\Psi_{I-1,I-1}
-4\Psi_{I,I}+4\Psi_{I,I+1}-\Psi_{I+1,I+1}-\Psi_{I,I+2}\right)\cr&&
\equiv \left[4
g^2\left(\sin^2\frac{p_1}{2}+\sin^2\frac{p_2}{2}\right)-\frac{g^4}{8}\left(\sin^4\frac{p_1}{2}+\sin^4\frac{p_2}{2}\right)\right]\Psi_{II}
\label{schr2II}\\\cr
&&\left(D_{1~loop}+D_{2~loop}\right)\circ\Psi_{I,I+1}=\cr
&&g^2\left(-\Psi_{I+1,I+1}-\Psi_{I-1,I+1}+4 \Psi_{I,I+1}
-\Psi_{I,I+2}-\Psi_{I,I}\right)\cr
&&+\frac{g^4}{2}\left(-\Psi_{I,I+3}+4\Psi_{I+1,I+1}+4 \Psi_{I,I+2}
-14\Psi_{I,I+1}+4\Psi_{I,I}+4\Psi_{I-1,I+1}-\Psi_{I-2,I+1}\right)\cr&&
\equiv \left[4
g^2\left(\sin^2\frac{p_1}{2}+\sin^2\frac{p_2}{2}\right)-\frac{g^4}{8}\left(\sin^4\frac{p_1}{2}+\sin^4\frac{p_2}{2}\right)\right]\Psi_{I,I+1}
\label{schr2II+1}
\end{eqnarray}
We regard these equations as determining $p_i$.

Using (\ref{2loopsansatz}) and (\ref{S0}) in (\ref{schr2II+1}) the
function $f_0(\m_1,\m_2)$ is uniquely derived as
\begin{equation}
f_0(\m_1,\m_2)=-\frac{(\m_1-1)(\m_2-1)(\m_1-\m_2)}{\m_2(1+\m_1(\m_2-2))}
\label{f0}
\end{equation}
Plugging (\ref{f0}) in (\ref{schr2II}) one can fix also the function
$S_1(\m_1,\m_2)$ as
\begin{equation}
 S_1(\m_2, \m_1)=-\frac{(\m_1-1)^2(\m_2-1)^2(\m_1-\m_2)(1+\m_1 \m_2)}
 {\m_2^2(1+\m_1(\m_2-2))^2}
\label{S1}
\end{equation}
Using (\ref{S0}) and (\ref{S1}) the Bethe equation (\ref{bethe2})
becomes
\begin{equation}\label{bethe2agree}
e^{i
p_2(kM+2)}=e^{i(p_1+p_2)}\left[\frac{\frac{1}{2}\cot\frac{p_2}{2}-\frac{1}{2}\cot\frac{p_1}{2}+i}
{\frac{1}{2}\cot\frac{p_2}{2}-\frac{1}{2}\cot\frac{p_1}{2}-i}-\frac{\l}{4\pi^2}\frac{\sin
p_1-\sin p_2}
{\left(\frac{1}{2}\cot\frac{p_2}{2}-\frac{1}{2}\cot\frac{p_1}{2}-i\right)^2}
\right]
\end{equation}
This is equivalent to Eq.~(\ref{bethetwoloops}) expanded to the
first order in $\l$. We have thus demonstrated that the PABA in
Eq.~(\ref{PABA}) solves the eigenvalue equations for the dilatation
operator in the form (\ref{D1IJ})-(\ref{D2II+1}) and that the
process of finding these solutions is equivalent to solving the
twisted Bethe equations for the $\mathcal{N}=2$ theory up to two
loops.

\section{Three loops}

The three loop operator dimensions cannot be gotten by direct
computation in Yang-Mills perturbation theory, or equivalently, by
the perturbative asymptotic Bethe ansatz approach that we used for
two loops in the previous Section. The reason is that, so far, no
explicit expression for the dilatation operator in terms of fields
and their derivatives is available at three loop order. Our approach
to computing at three loops will therefore be to assume that the
twisted Bethe ansatz, summarized in Eqs.~(\ref{tb1})-(\ref{tb3}),
correctly describes the spectrum and to derive the three-loop
correction to operator dimensions from it.

For this purpose we have to keep $O(\lambda^2)$ terms in
Eq.(\ref{tb1}) so that the twisted Bethe equation now reads
\begin{eqnarray}\label{threeloopstwisted}&&e^{i
p_2(k M+2)}=e^{i(p_1+p_2)}\cr
&&\!\!\!\!\!\frac{\frac{1}{2}\cot\frac{p_2}{2}+\frac{\l}{8\pi^2}\sin
p_2+\frac{\l^2}{64\pi^4}\sin p_2(\cos
p_2-1)-\frac{1}{2}\cot\frac{p_1}{2}-\frac{\l}{8\pi^2}\sin
p_1-\frac{\l^2}{64\pi^4}\sin p_1(\cos
p_1-1)+i}{\frac{1}{2}\cot\frac{p_2}{2}+\frac{\l}{8\pi^2}\sin
p_2+\frac{\l^2}{64\pi^4}\sin p_2(\cos
p_2-1)-\frac{1}{2}\cot\frac{p_1}{2}-\frac{\l}{8\pi^2}\sin
p_1-\frac{\l^2}{64\pi^4}\sin p_1(\cos p_1-1)-i}\cr&&
\end{eqnarray}
We look for a solution of this equation by means of momenta of the
following form
\begin{eqnarray}
\label{threeloopsansatz} p_1=\frac{2 n_1\pi}{k
M}+\frac{A\pi}{M^2}+\lambda'\frac{B\pi}{M^2}+\lambda'^2\frac{C\pi}{M^2}
\cr p_2=\frac{2 n_2\pi}{k
M}-\frac{A\pi}{M^2}-\lambda'\frac{B\pi}{M^2}-\lambda'^2\frac{C\pi}{M^2},
\end{eqnarray}
where $A$ and $B$ have been computed at lower loops, Eqs. (\ref{A})
and (\ref{B}). Recall that $\lambda' = {\small\frac{\lambda}{M^2}}$.
Requiring that the Bethe equations are satisfied at order
$\lambda'^2$ we fix $C$ as
\begin{equation}\label{C} C=\frac{n_1^2 n_2^2\left(n_1^2-n_1
n_2+n_2^2\right)}{2 k^6(n_2-n_1)}
\end{equation}

The eigenvalue formula eq.(\ref{tb3}) expanded up to three loops
gives
\begin{eqnarray}
\label{threeloopsanomalousdimension} & \Delta= & k M+2+
\frac{\lambda'
M^2}{2\pi^2}\left(\sin^2{\frac{p_1}{2}}+\sin^2{\frac{p_2}{2}}\right)-\frac{\lambda'^2
M^4}{8
\pi^4}\left(\sin^4{\frac{p_1}{2}}+\sin^4{\frac{p_1}{2}}\right)\cr &&
+\frac{\lambda'^3 M^6}{16
\pi^6}\left(\sin^6{\frac{p_1}{2}}+\sin^6{\frac{p_2}{2}}\right)+
O(\l'^4)
\end{eqnarray}
Taking into account the $\l'$ dependence of the momenta given in
(\ref{threeloopsansatz}) and expanding in $\lambda'$ and
$\frac{1}{M}$, we obtain the planar three loop result up to the
first finite size correction
\begin{equation}\label{threeloopsanomalousdimension2}
\Delta_{3~loops}=\frac{\lambda'^3}{16}\left[\frac{n_1^6+n_2^6}{k^6}-
\left(\frac{2}{k M}\right)\frac{3n_1^6+3n_1^5
n_2+4n_1^3n_2^3+3n_1n_2^5+3n_2^6}{k^6}+O\left(\frac{1}{M^2}\right)\right]\
.
\end{equation}
This result has to be compared with the $1/R^2$ corrections to the
pp-wave energy spectrum of the corresponding string states.

As a consistency check, we see that when we set the wrapping number
to zero to get the $\mathcal{N}=4$ state, i.e. put  $n_2=-n_1$, it
provides the $\mathcal{N}=4$ result, in beautiful agreement with the
one quoted in Refs.~\cite{Serban:2004jf}, \cite{Beisert:2004hm}.

\section{On the string side of the duality}

In the previous Sections, we discussed the expansion to leading
order in $\frac{1}{M}$ about the MRV limit of the $\mathcal{N}=2$
quiver gauge theory.  The string dual to the quiver gauge theory is
the IIB superstring on the $AdS_5\times S^5/Z_M$ background. The MRV
limit of the $\mathcal{N}=2$ theory corresponds to the simultaneous
Penrose limit and large $M$ limit of the $AdS_5\times S^5/Z_M$
orbifold where the ratio $R_-= \frac{R^2}{2M}$ is held constant.
Here, $R$ is the radius of curvature of $AdS_5\times S^5/Z_M$. The
result is the pp-wave background where the null coordinate has been
periodically identified with radius $R_-$. String theory in that
background is described by a DLCQ version of the string theory on
the maximally symmetric pp-wave.  The $\frac{1}{M}$ expansion of
Yang-Mills theory about the MRV limit corresponds to an expansion in
the ratio $ \frac{1}{M}=\frac{2R_-}{R^2}$ about the pp-wave
space-time.

 Corrections of this kind have already been analyzed in some detail
for the case of $\mathcal{N}=4$ super Yang-Mills theory -- string on
$AdS_5\times S^5$ duality in Ref.~\cite{Callan:2003xr}.  They
considered the leading correction to the BMN limit, which was an
expansion in the inverse R-charge $\frac{1}{J}$ of Yang-Mills theory
or $\frac{\alpha'}{R^2}$ in string theory. In this section, we will
generalize their computation to the case of the DLCQ string on the
pp-wave background.  We will compare the result with our
computations of $1/M$-corrections in the quiver gauge theory.

The exact spectrum of states of the string theory on the pp-wave
background, as well as the DLCQ of the pp-wave background are
well-known.  Our goal is to find corrections to the energies of
these states to order $ \frac{2R_-}{R^2}$. The technique to be used
is to first find the correction to the string sigma model which
arises from an expansion of the space-time metric and other
background fields about the pp-wave. This yields an interaction
Hamiltonian. The strategy is then to compute corrections to the
energy spectrum by evaluating matrix elements of this interaction
Hamiltonian in the pp-wave string theory states.  The coefficient of
the interaction Hamiltonian contains the factor $ \frac{2R_-}{R^2}$.

In the case of $AdS_5\times S^5$ background, the terms in the
interaction Hamiltonian which contain two bosonic creation and two
bosonic annihilation operators are expressed in terms of the string
oscillators as~\cite{Callan:2003xr}
\begin{eqnarray}
\label{Hcorrected} H_{BB} & = &
    -\frac{1}{32 p^+ R^2}\sum \frac{\delta(n+m+l+p)}{\xi}
    \times
\nn\\
& & \biggl\{
    2 \biggl[ \xi^2
    - (1 - k_l k_p k_n k_m )
     +  \omega_n \omega_m k_l k_p
      +  \omega_l \omega_p k_n k_m
    + 2 \omega_n \omega_l k_m k_p
\nn\\
& &     + 2 \omega_m \omega_p k_n k_l
    \biggr]
    a_{-n}^{\dagger A}a_{-m}^{\dagger A}a_l^B a_p^B
   +4 \biggl[ \xi^2
    - (1 - k_l k_p k_n k_m )
     - 2 \omega_n \omega_m k_l k_p
     +  \omega_l \omega_m k_n k_p
\nn\\
& &   -  \omega_n \omega_l k_m k_p
    -  \omega_m \omega_p k_n k_l
    + \omega_n \omega_p k_m k_l \biggr]
    a_{-n}^{\dagger A}a_{-l}^{\dagger B}a_m^A a_p^B
     + 4  \biggl[8 k_l k_p
    a_{-n}^{\dagger i}a_{-l}^{\dagger j}a_m^i a_p^j
\nn\\
& &     + 2 (k_l k_p +k_n k_m)
    a_{-n}^{\dagger i}a_{-m}^{\dagger i}a_l^j a_p^j
    +(\omega_l \omega_p+ k_l k_p -\omega_n
    \omega_m- k_n k_m)a_{-n}^{\dagger i}a_{-m}^{\dagger i}a_l^{j'} a_p^{j'}
\nn\\
& &     -4 ( \omega_l \omega_p- k_l k_p)
    a_{-n}^{\dagger i}a_{-l}^{\dagger j'}a_m^i a_p^{j'}
    -(i,j \rightleftharpoons i',j')
    \biggr]\biggr\} ,
\end{eqnarray}
where $p^+$ is the space-time momentum conjugate to the light-cone
coordinate $x^-$, $\xi \equiv \sqrt{\omega_n \omega_m \omega_l
\omega_p}$ , $\omega_n=\sqrt{1+k_n^2}$ and
$k_n^2=\frac{n^2}{\alpha^{\prime 2} {p^+}^2}=\lambda^\prime n^2$,
with $\lambda^\prime=g^2_{YM}N/J^2$. The indices $l,m,n,p$ run from
$-\infty$ to $+\infty$. The presence of the R-R flux breaks the
transverse $SO(8)$ symmetry of the metric to $SO(4) \times SO(4)$.
Consequently the notation distinguishes sums over indices of the
transverse coordinates in the first $SO(4)$ ($i,j,..$), the second
$SO(4)$ ($i^\prime,j^\prime,..$) and over the full $SO(8)$
($A,B,..$). The operators in (\ref{Hcorrected}) are in a
normal-ordered form. Since $H_{BB}$ was derived as a classical
object, the correct ordering on the operators is not defined  and
the ambiguity thus arising can be kept into account by introducing a
normal ordering function $N_{BB}(k_n^2)$.  Such normal-ordering
function can however be set to zero following the prescription of
Ref.\cite{Callan:2003xr}.

The DLCQ version of (\ref{Hcorrected}) can be obtained by taking
into account that the light-cone momentum $p^+$ along the
compactified light-cone direction ($x^-\sim x^-+2\pi R^-$) is
quantized as $p^+=k/(2 R_-)$. $R_-$ is related to $R$ through
$R_-=R^2/(2 M)$ so that $p^+=k M/ R^2$ and $R^2=\sqrt{4 \pi
g_s\alpha^{\prime 2} N M}$. The Yang-Mills theory coupling constant
is then identified with the superstring coupling constant $g_s$ in
the usual way $4\pi g_s=g^2_{\rm YM}$ and the double scaling limit
is realized by sending both $N$ and $M$ to infinity and keeping the
ratio $N/M$ fixed, so that
$R_-=\frac{\alpha'}{2}\sqrt{g^2_{YM}\frac{N}{M}}=\frac{\a'}{2}\sqrt{\l'}$
is also held fixed. As noticed in the introduction, the definition
of $\lambda'$ is in this case related to the $YM$ coupling constant
through an analogue of the usual definition $\frac{1}{(\a'
p^+)^2}=\frac{g^2_{YM} NM}{(kM)^2}\equiv \frac{\lambda'}{k^2}$. This
gives for the frequencies $\omega_n$ in (\ref{Hcorrected}) the
formula $\omega_n=\sqrt{1+\lambda'\frac{n^2}{k^2}}$.

In the case of the $\mathcal{N}=2$ operator (\ref{bigoperator}), the
dual string state is the symmetric traceless two-impurity state
created by the action of the following combination of bosonic
creation operators on the string vacuum\footnote{We use the notation
of Ref.~\cite{Callan:2003xr}, where the representations of
$SO(4)\times SO(4)$ are classified using an $SU(2)$ notation as
$SO(4)\approx SU(2)\times SU(2)$.}
\begin{equation} \label{ST}
|[{\bf 1},{\bf 1};{\bf 3},{\bf 3}]>=\left[a_{n_1}^{\dagger a}
a_{n_2}^{\dagger b}+a_{n_1}^{\dagger b} a_{n_2}^{\dagger
a}-\frac{1}{2}\delta^{ab}a_{n_1}^{\dagger g} a_{n_2}^{\dagger
g}\right]|0\rangle
\end{equation}
where $n_1+n_2=k~\ell$.

The general matrix elements of the DLCQ version $H_{BB}^{Z_M}$ of
(\ref{Hcorrected}) between space-time bosons built out of bosonic
string oscillators have the following explicit form
\begin{eqnarray}
\label{matrixelements} &&\langle 0|\, a_{-n_2}^A
a^B_{-n_1}\,H^{Z_M}_{BB}\,a_{n_1}^{\dagger C} a_{n_2}^{\dagger
D}\,|0\rangle =-\frac{1}{2 R^2p^+}
\frac{1}{\sqrt{1+\lambda'\frac{n_1^2}{k^2}}\sqrt{1+\lambda'\frac{n_2^2}{k^2}
}}\cr &&\left\{\delta^{AB}\delta^{CD}\lambda'
\left[\frac{n_1^2}{k^2}+\frac{n_2^2}{k^2}+2\lambda'\frac{n_1^2
n_2^2}{k^4} +2 \frac{n_1 n_2}{k^2}\sqrt{1+\lambda'\frac{n_1^2}{k^2}
}\sqrt{1+\lambda'\frac{ n_2^2}{k^2}}\right]\right.\cr
&&\left.+\delta^{AC}\delta^{BD}\lambda'\left[\frac{n_1^2}{k^2}+\frac{n_2^2}{k^2}
+2\lambda'\frac{n_1^2 n_2^2}{k^4} -2 \frac{n_1
n_2}{k^2}\sqrt{1+\lambda'\frac{n_1^2}{k^2} }\sqrt{1+\lambda'\frac{
n_2^2}{k^2}}\right]\right.\cr &&\left.+\lambda' \left[2 \frac{n_1
n_2}{k^2}\left(\delta^{ab}\delta^{cd}+\delta^{ac}\delta^{bd}\right)+\frac{(n_1^2+
n_2^2)}{k^2}\delta^{ad}\delta^{bc}\right]\right.\cr
&&\left.-\lambda'\left[2 \frac{n_1 n_2}{k^2}\left(\delta^{a^\prime
b^\prime}\delta^{c^\prime d^\prime}+\delta^{a^\prime
c^\prime}\delta^{b^\prime d^\prime}\right)+\frac{(n_1^2+
n_2^2)}{k^2}\delta^{a^\prime d^\prime}\delta^{b^\prime
c^\prime}\right]\right\}
\end{eqnarray}
where lower-case $SO(4)$ indices $a,b,c,d\in 1,\dots ,4$ mean that
the corresponding $SO(8)$ labels $A,B,C,D$ all lie in the first
$SO(4)$, while the indices $a',b',c',d'\in 5,\dots ,8$ mean that the
$SO(8)$ labels lie in the second $SO(4)$ $(A,B,C,D \in 5,\dots ,8)$.

Eq. (\ref{matrixelements}) can be used to evaluate the first order
correction to the energy of the state (\ref{ST}), namely the matrix
element $<[{\bf 1},{\bf 1};{\bf 3},{\bf 3}]|H_{BB}^{Z_M}|[{\bf
1},{\bf 1};{\bf 3},{\bf 3}]>$. Summing all the contributes and
dividing the result by the norm of the state $$<[{\bf 1},{\bf
1};{\bf 3},{\bf 3}]|[{\bf 1},{\bf 1};{\bf 3},{\bf
3}]>=2(1+\frac{1}{2}\delta^{ab})$$ one gets the desired first
curvature correction to the spectrum of the states (\ref{ST}). The
final result for the energy levels for a two impurity state with
discrete light-cone momentum $k$, exact to all orders in
$\lambda^\prime$, is
\begin{eqnarray}
\label{finenergy}&& E(n_1,n_2)=
\sqrt{1+\lambda'\left(\frac{n_1}{k}\right)^2
}+\sqrt{1+\lambda'\left(\frac{n_2}{k}\right)^2}\cr
&&-\frac{\lambda'}{k
M}\left[\frac{\frac{n_1^2}{k^2}+\frac{n_2^2}{k^2}+\lambda'\frac{
n_1^2n_2^2}{k^4}+\frac{n_1 n_2}{k^2}-\frac{n_1
n_2}{k^2}\sqrt{1+\lambda'\left(\frac{n_1}{k}\right)^2
}\sqrt{1+\lambda'\left(\frac{n_2}{k}\right)^2
}}{\sqrt{1+\lambda'\left(\frac{n_1}{k}\right)^2
}\sqrt{1+\lambda'\left(\frac{n_2}{k}\right)^2
}}\right]+O\left(\frac{1}{M^2}\right)\cr&&
\end{eqnarray}
where the small parameter governing the strength of the perturbation
has been converted from $1/(R^2p^+ )$ to $1/(k M)$ in order to make
the comparison with the finite size corrections of the gauge theory
results more clear. Notice that for $n_1=-n_2$ (\ref{finenergy})
gives back the ${\mathcal N}=4$ result of Ref.\cite{Callan:2003xr},
as it should.

A $\lambda^\prime$ expansion of (\ref{finenergy}) up to
$O(\lambda'^2)$ shows perfect agreement with the gauge theory
calculations at one and two loops, Eqs.(\ref{finitesizeoneloop}) and
(\ref{twoloopsanomalousdimension}). As for the parent
$\mathcal{N}=4$ theory~\cite{Serban:2004jf, Beisert:2004hm}, the
disagreement between the two sides of the duality is manifest at
three loops, where the finite size correction to the string energy
\begin{eqnarray}
\label{finitesizethreeloops} E_{3~loops}=
\frac{\lambda'^3}{16}&&\left[\frac{n_1^6+n_2^6}{k^6}
-\left(\frac{2}{k M}\right)\frac{3n_1^6+3n_1^5 n_2+n_1^4
n_2^2+2n_1^3 n_2^3+n_1^2 n_2^4+3 n_1 n_2^5+3 n_2^6}{k^6}\right.
\cr&&\left.+~O\left(\frac{1}{M^2}\right)\right]
\end{eqnarray}
does not match its gauge dual result
(\ref{threeloopsanomalousdimension2}).

\section{The S-matrix dressing factor}

Integrable structures have been found also in the $AdS_5\times S^5$
string sigma model: from a classical point of view integral Bethe
equations were derived in the thermodynamic
limit~\cite{Kazakov:2004qf}, while quantum corrections are believed
to yield discrete equations describing a finite number of
excitations.

The agreement between the anomalous dimensions of the
$\mathcal{N}=4$ gauge theory operators in the near-BMN limit and the
string energies in the near-plane wave limit up to two gauge theory
loops suggests that, if we wish to describe the string excitations
by the language of a spin chain, the string dynamics should be given
by the BDS chain.

The three loop disagreement can actually be encoded by ``dressing''
the gauge theory S-matrix ($i.e.$ the r.h.s. of the Bethe equations
for the BDS chain) by a multiplicative factor. From these equations
one derives a solution for the momenta of the string excitations
which plugged in the BDS dispersion relation (\ref{tb3}) reproduce
the near-plane wave string energies, both in the thermodynamic limit
and in the few impurity case~\cite{Callan:2004ev, Arutyunov:2004xr}.

The near-plane wave string energies can therefore be computed in the
$AdS_5\times S^5$ IIB superstring theory by the following Bethe
equations:
\begin{equation} \label{Bethestring}
e^{i p_{j}L}=\prod_{l=1~;~l\neq j}^{\cal M} S_{string}(p_j,p_l),
\end{equation}
with $L=J+\cal M$ and
\begin{equation} \label{stringSparent}
S_{string}(p_j,p_l)=
\frac{\varphi_{j}-\varphi_{l}+i}{\varphi_{j}-\varphi_{l}-i}
\exp\left\{2i \sum_{r=0}^{\infty}\Big(\frac{\lambda}{16
\pi^2}\Big)^{r+2}[q_{r+2}(p_j)q_{r+3}(p_l)-q_{r+2}(p_l)q_{r+3}(p_j)]\right\}
\end{equation}
where the BDS rapidities are defined in (\ref{b2}) and the
exponential term is the so called dressing factor, expressed as a
function of the BDS conserved charges
\begin{equation}
\label{dispersionlaws}
q_r(p_j)=\frac{2\sin{(\frac{r-1}{2}p_j})}{r-1}\left(\frac{\sqrt{1+\frac{\lambda
}{\pi^2}\sin^2{\frac{p_j}{2}}}-1}{\frac{\lambda}{4
\pi^2}\sin{\frac{p_j}{2}}}\right)^{r-1}
\end{equation}

In particular, the second charge $q_2(p_j)$ is the energy of a
single excitation and the energy of a string state with ${\cal M}$
excitations is given by
\begin{equation}
E=\frac{\lambda}{8 \pi^2}\sum_{j=1}^{\cal M}q_2(p_j) \label{stren}
\end{equation}

We will now discuss the two magnon case in the orbifolded theory and
show that the same dressing factor allows one to compute the DLCQ
string energies by means of a Bethe ansatz. The two magnon
scattering however is not as trivial as in the parent theory, since
the excitations are not forced by the level matching condition to
carry opposite momenta.

It is not difficult to check that  the string spectrum
(\ref{finenergy}) coincides with (\ref{stren}) up to $O(\lambda'^3)$
 with $\mathcal{M}=2$ if the magnon momenta have the form
\begin{eqnarray}
\label{stringmomenta} p_1=\frac{2 n_1\pi}{k
M}+\frac{A\pi}{M^2}+\lambda'\frac{B\pi}{M^2}+\lambda'^2\frac{C'\pi}{M^2}
\cr p_2=\frac{2 n_2\pi}{k
M}-\frac{A\pi}{M^2}-\lambda'\frac{B\pi}{M^2}-\lambda'^2\frac{C'\pi}{M^2},
\end{eqnarray}
with the same $A$ and $B$ found in the gauge theory, Eqs. (\ref{A})
(\ref{B}), and  $C'$ given by
\begin{equation}\label{Cstring} C'=\frac{n_1^2 n_2^2\left(n_1^2+n_2^2\right)}{4 k^6(n_1-n_2)}
\end{equation}

We conjecture that the string $S$-matrix for the
AdS$_5\times$S$^5/Z_M$ IIB superstring is given by
(\ref{stringSparent}) with the addition of a twist factor which
coincides with the one used in the gauge theory
\begin{eqnarray} \label{stringSorb}
&&S^{orb.}_{string}(p_j,p_l)=\omega^l
\frac{\varphi_{j}-\varphi_{l}+i}{\varphi_{j}-\varphi_{l}-i}  \cr & &
\exp{\bigg(2i \sum_{r=0}^{\infty}\Big(\frac{\lambda}{16
\pi^2}\Big)^{r+2}[q_{r+2}(p_j)q_{r+3}(p_l)-q_{r+2}(p_l)q_{r+3}(p_j)]\bigg)}
\end{eqnarray}
with $\omega^l=e^{i(p_1+p_2)}$ for the two magnon case. It is easy
to see that the Bethe equations
\begin{equation} \label{Bethestring}
e^{i p_{2}(k M+2)}= S^{orb.}_{string}(p_2,p_1),
\end{equation}
are in fact satisfied if $p_1$ and $p_2$ are exactly
(\ref{stringmomenta}), with the constants $A$, $B$ and $C$ given in
(\ref{A}), (\ref{B}) and (\ref{Cstring}).

Thus we have proved that the dressing factor for the orbifolded
theory equals that of the parent theory and therefore, as for the
gauge theory, the spectrum can be obtained by just twisting the
parent Bethe equations: the three loop disagreement is inherited and
does not depend on the orbifold projection.

\section{Summary}

In this Paper, we have computed the first finite size correction to
the anomalous dimension of two-impurity states about the double
scaling limit of the $\mathcal{N}=2$ quiver gauge theory and the
analogous quantity in the IIB superstring propagating on the
plane-wave background with a periodically identified null
coordinate.

In the gauge theory  the anomalous dimensions are computed by two
independent techniques that agree with each other. We have solved,
up to three loops and the first finite size correction, the twisted
Bethe equations conjectured in Ref.~\cite{Beisert:2005he} for the
orbifolded theory. Then we have provided an ansatz for the
eigenstate of the dilatation operator that up to two loops gives the
same spectrum derived with the other procedure. The eigenvalue
equation for this wave function reduces to the twisted Bethe
equation.

On the string theory side the computation is done by evaluating the
first curvature correction to the pp-wave DLCQ spectrum of a bosonic
two excitation state.

We have found that the gauge theory and the string theory results
agree up to two loop order, but there is a disagreement at three
loops. This disagreement is similar to, and a slight generalization
of the one which is known to exist at three loop order in the
analogous computation in $\mathcal{N}=4$ super Yang-Mills theory
expanded about the BMN limit~\cite{Serban:2004jf, Beisert:2004hm}.

In Summary, the results of this Paper are

\begin{eqnarray}
\Delta_{YM}&=& kM+2 +\frac{\lambda'}{2}\left[
\frac{n_1^2+n_2^2}{k^2}
\right]-\frac{\lambda'^2}{8}\left[\frac{n_1^4+n_2^4}{k^4} \right]+
\frac{\lambda'^3}{16}\left[\frac{n_1^6+n_2^6}{k^6}  \right]+...
\nonumber \\ &+& \frac{\lambda'}{kM} \left[  -
\frac{\left(n_1^2+n_2^2\right)}{k^2} +\frac{\lambda'}{2}
\frac{n_1^4+n_1^3 n_2+n_1 n_2^3+n_2^4}{k^4}  \right. \nonumber \\
&-&\left. \frac{\lambda'^2}{8}
  \frac{3n_1^6+3n_1^5
n_2+4n_1^3n_2^3+3n_1n_2^5+3n_2^6}{k^6}+...\right]
\end{eqnarray}

\begin{eqnarray}
\Delta_{string} &=& kM+2 +\frac{\lambda'}{2}\left[
\frac{n_1^2+n_2^2}{k^2}
\right]-\frac{\lambda'^2}{8}\left[\frac{n_1^4+n_2^4}{k^4} \right]+
\frac{\lambda'^3}{16}\left[\frac{n_1^6+n_2^6}{k^6}  \right]+...
\nonumber \\ &+& \frac{\lambda'}{kM} \left[  -
\frac{\left(n_1^2+n_2^2\right)}{k^2} +\frac{\lambda'}{2}
\frac{n_1^4+n_1^3 n_2+n_1 n_2^3+n_2^4}{k^4}  \right. \nonumber \\
&-&\left. \frac{\lambda'^2}{8} \frac{3n_1^6+3n_1^5 n_2+n_1^4
n_2^2+2n_1^3 n_2^3+n_1^2 n_2^4+3 n_1 n_2^5+3 n_2^6}{k^6}+... \right]
\end{eqnarray}

The first two lines of each of the above expressions are identical
and they differ in the third line.

We have finally shown that the DLCQ string spectrum is obtained by
twisting the string Bethe ansatz proposed in
Ref.~\cite{Arutyunov:2004xr}. The three loop disagreement is encoded
in a ``dressing factor'' added to the gauge theory S-matrix, which
coincides with the one of the $\mathcal{N}=4$ theory.

Our computations are consistent with integrability of
$\mathcal{N}=2$ quiver gauge theory in the MRV limit and its string
theory dual, DLCQ type IIB superstring theory on a plane wave
background with a compactified null direction.

\section*{Acknowledgements:} The work of
D.~Astolfi, V.~Forini and G.~Grignani is supported in part by the
I.N.F.N.~ and M.I.U.R.~of Italy and by the PRIN project 2005-024045
``Symmetries of the Universe and of the Fundamental Interactions''.
V.~Forini and G.~Grignani acknowledge the hospitality of the Pacific
Institute for Theoretical Physics and the University of British
Columbia where parts of this work were done. The work of
G.W.~Semenoff is supported by the Natural Sciences and Engineering
Research Council of Canada.  G.W.~Semenoff acknowledges the
hospitality of the University of Perugia where some of this work was
done.

\end{document}